\documentclass[twocolumn,aps,prb,showpacs,preprintnumbers,amsmath,amssymb, superscriptaddress]{revtex4}

\usepackage{graphicx}
\usepackage{dcolumn}
\usepackage{bm}

\begin{document}


\title{Critical behavior of diluted magnetic semiconductors: the apparent violation and 
the eventual restoration of the Harris 
criterion for all regimes of disorder}


\author{D. J. Priour, Jr}
\affiliation{Department of Physics, University of Missouri, Kansas City, MO 64110, USA}
\author{S. Das Sarma}
\affiliation{Condensed Matter Theory Center, Department of Physics, University of Maryland, College Park, MD 20742-4111, USA}


\date{\today}

\begin{abstract}
Using large-scale Monte Carlo calculations, we 
consider strongly disordered Heisenberg models on a cubic lattice with missing sites (as in 
diluted magnetic semiconductors such as $\textrm{Ga}_{1-x}\textrm{Mn}_{x}\textrm{As}$).  For disorder
ranging from weak to strong levels of dilution,
we identify Curie temperatures and 
calculate the critical exponents $\nu$, $\gamma$, $\eta$, and $\beta$ finding, per the Harris criterion, 
good agreement with critical indices for the pure Heisenberg model where there is no disorder component. 
Moreover, we find that thermodynamic quantities (e.g. the second moment of the magnetization per spin) 
self average at the ferromagnetic transition temperature with 
relative fluctuations tending to zero with increasing system size.
We directly calculate effective critical exponents for $T > T_{c}$, yielding values 
which may differ significantly from the critical indices for the pure system, 
especially in the presence of strong disorder.  Ultimately,
the difference is only apparent, and eventually disappears when $T$ is very close to $T_{c}$.   
\end{abstract}

\maketitle


\section{Introduction}

Technologically relevant magnetic materials such as diluted magnetic semiconductors (DMS) are 
characteristically strongly disordered due to the low concentration of random magnetic moments (e.g.     
$\textrm{Ga}_{1-x}\textrm{Mn}_{x}As$ where 5\% - 12 \% of the Ga sites are occupied by substituent Mn ions).
DMS materials such as $\textrm{Ga}_{1-x}\textrm{Mn}_{x} \textrm{As}$ have been modeled theoretically 
using a classical Heisenberg model on an fcc lattice where the Hamiltonian is $\mathcal{H} = 
\sum_{i,j}J_{ij} \mathbf{S}_{i} \cdot \mathbf{S}_{j}$ with $J(r_{ij})$ being a carrier (hole) mediated 
random indirect exchange coupling between moments separated by a distance $r_{ij}$ given by 
$J(r) = J_{0} e^{-r/l} r^{-4} [ \sin ( 2 k_{\mathrm{F}} r) - 2 k_{\mathrm{F}} r \cos (2 k_{\mathrm{F}} r)]$.
$k_{\mathrm{F}} = (\frac{3}{2} \pi^{2} n_{c} )^{1/3}$ is the Fermi wave number, $n_{c}$ is the hole 
density, and $l$ is the damping scale.

While individual parameters such as the ferromagnetic transition temperature $T_{c}$ have been calculated in 
theoretical studies~\cite{DJP1,DJP2}, the critical behavior of strongly disordered Heisenberg models on a three dimensional
lattice has not been understood in detail in the context of a direct numerical calculation.
At the ferromagnetic transition, thermodynamic quantities scale as power laws in the reduced 
temperature $t = (T - T_{c})/T_{c}$ with, e.g., the magnetization varying as $m \propto t^{\beta}$, the
correlation length scaling as $\xi \propto t^{-\nu}$, and $\chi \propto t^{-\gamma}$ for 
the magnetic susceptibility; hence
critical exponents such as $\beta$, $\nu$, and $\gamma$ (up to prefactors specific to the 
model under consideration) completely specify the critical behavior near $T_{c}$ where $t \ll 1$.

Our task is to determine the extent to which the critical behavior of the three dimensional Heisenberg model
is influenced by disorder (in the form of randomly removed magnetic 
moments), and we have found the most singular contributions to critical behavior to 
be unaffected by disorder whether only a few magnetic moments are removed or 
the majority of magnetic impurities are missing in cases of strong disorder.    
A theoretical result (derived from a renormalization group 
calculation) known as the Harris criterion~\cite{Harris} holds that the sign of the specific heat exponent $\alpha$ determines 
whether the critical exponents are altered.  Specifically, although modifications in the universality class are   
expected for $\alpha > 0$, the Harris criterion predicts that disorder will not affect the critical 
exponents when $\alpha < 0$.  The hyper-scaling identity $\alpha = 2 - d \nu$ implies that the condition for stable 
critical behavior is $ \nu > 2/d$, $d = 3$ being the dimensionality of our system.  In particular, since $ \nu = 0.714 
> 0.67$ for the Heisenberg model~\cite{Peli}, the Harris result precludes disorder induced shifts in the critical exponents.  
With careful finite size scaling analysis, we have indeed confirmed that critical behavior in the disordered models 
conforms to the 3D Heisenberg universality class. 
An important finding of our detailed numerical study is, however, the fact 
that the effective critical exponents of the strongly disordered model may very well manifest an 
apparent violation of the Harris criterion (i.e. a deviation from the corresponding pure Heisenberg model values) 
away from the critical temperature, thus possibly considerably complicating the interpretation of experimental data."

The results of our numerical calculations are consistent with experiment where 
the local critical behavior of thermodynamic quantities such as the magnetic susceptibility 
$\chi$ (e.g. the slope $\gamma_{\mathrm{eff}} =  d \log (\chi )/d \log (t)$ of the log-log plot
in the case of the magnetic susceptibility) 
differ from the critical indices of the pure case with $c = 1.0$ for 
intermediate values of the reduced temperature. 
Ultimately, the effective critical exponents converge for sufficiently small $t$ to   
the critical behavior of the model with no disorder.
Similarly, we examine finite size systems, and we would obtain results for 
critical behavior which differ from those of the pure model
if we extrapolate to the bulk limit in na\"{i}ve manner.  However, by taking into account
corrections to scaling, we compensate for finite size effects and obtain critical 
exponents identical to those of the pure Heisenberg model.

Using large-scale Monte Carlo simulations, we calculate critical exponents for 
the disordered Heisenberg model on a 3D lattice.   
Hence, we show that the universality class remains unaltered from regimes where the model is weakly disordered and
only a few magnetic moments are removed to cases such as $c = 0.4$ (the site percolation threshold for the simple
cubic lattice is $c = 0.3116$ where on average fewer than half of the magnetic ions
participate in a ferromagnetically ordered phase).

Another component of the Harris criterion is the prediction that thermodynamic variables such as the magnetization 
$m$ and magnetic susceptibility $\chi$ do (do not ) self-average at $T_{c}$ in the bulk limit when $\nu > (<) 2/3$.
The extent of self-averaging may be quantified via the parameter $g_{2} = ([ \langle m^{2} \rangle^{2} ] 
- [ \langle m^{2} \rangle]^{2})/[ \langle m^{2} \rangle ]^{2}$~\cite{g2}, the relative 
variance of $[ \langle m^{2} \rangle ]$ with respect to disorder where $m$ is the magnetization, angular brackets 
indicate thermal averages, and square brackets refer to disorder averaging.
For the Heisenberg model, we find self-averaging to be intact with $g_{2}$ ultimately decreasing 
after reaching a maximum for moderate sized systems containing on the order of a few hundred magnetic 
impurities.

In Section II, we discuss details of our numerical techniques for determining critical 
behavior of the disordered Heisenberg model.  Subtleties include the need for a careful calculation 
of the Curie Temperature $T_{c}$, and taking into account corrections to scaling which would otherwise 
lead to the conclusion that disorder has affected the critical behavior of the Heisenberg model; we find
that the universality class is not influenced by disorder, being identical to that of the pure model.

In Section III, we give results in tabular form for the critical exponents obtained in our calculation.
Explicit numerical values are given for the critical indices $\nu$, $\beta$, $\gamma$, and $\eta$ for 
disorder ranging from very weak (e.g. $c = 0.95$) to quite strong (i.e. $c = 0.4$).  In each 
case, we also provide the corresponding critical exponent (calculated by us) for the pure model, which is consistent with
the best and most recent values given in the literature.  

In Section IV, we provide the apparent critical exponents which differ from those of the pure model, 
and would be obtained for system sizes that are not 
sufficiently large.  Similarly, if one is not close enough to $T_{c}$ in experiment (generally, the reduced temperature 
$t = (T - T_{c})/T_{C}$ should be less than $10^{-3}$ to obtain the critical exponents of the pure Heisenberg model 
in systems with disorder), spurious apparent critical indices will be measured.
This apparent violation of the Harris criterion, even very slightly away from the critical temperature, is an 
important cautionary remark following directly from our Monte Carlo studies of the disordered model.

Finally, in the Appendix (Section V), we provide the Monte Carlo numerical results for 
thermodynamic variables such as the magnetization $m$ and magnetic susceptibility $\chi$.  
Also included are the corresponding theoretical results taking into account leading singular 
terms, as well as the first correction to scaling.  There is very good agreement between the Monte Carlo 
data and the results of the theoretical model (i.e. generally at least one part in $10^{3}$ or better).    

\section{Methods and Techniques in the Numerical Analysis}

Singularities in variables such as the specific heat and magnetic susceptibility are smoothened as $t \rightarrow 0$
and the correlation length $\xi$ becomes comparable to the system size $L$.  However, we can determine 
critical exponents by exploiting finite size scaling at $T_{c}$; the magnetization scales at $m \propto L^{-\beta/L}$, 
the thermal derivative $d \xi/dT$ of the correlation length $\xi$ varies as $d \xi/dT \propto L^{1/\nu}$, and the  
magnetic susceptibility $\chi$ diverges with increasing system size $L$ with
the singular dependence $\chi = c L^{3} ([\langle m^{2} \rangle] - [ \langle m \rangle ]^{2}) \propto L^{\gamma/\nu}$.
The critical exponents $\nu$, $\beta/\nu$, and $\gamma/\nu$ 
are obtained by calculating the appropriate thermodynamic quantities for many different system sizes
and carefully extrapolating to the thermodynamic limit.  Having calculated $\nu$, $\gamma$, and $\beta$, one
may obtain additional critical exponents such with the aid of hyperscaling relations.
As an example, the exponent $\eta$, given in terms of $\gamma$ and $\nu$ by $\eta = 2 - \gamma/\nu$,
is useful because it is a more sensitive parameter than $\gamma$ alone in gauging the universality class of a specific 
model.

To obtain critical exponents accurately, it is essential that calculations be performed as close as  
possible to $T_{c}$~\cite{Chen} since the temperature range where finite size scaling holds becomes narrower with increasing 
system size $L$.  To obtain the ferromagnetic transition temperature $T_{c}$ as precisely as possible, we 
numerically calculate the normalized correlation length $\xi/L$ following reference~\cite{Kim}.  For temperatures below 
$T_{c}$, $\xi/L$ ultimately increases with increasing $L$, while above the Curie temperature
$\xi/L$ eventually decreases.  We find $T_{c}$ by insisting that $\xi/L$ tend to a constant value
for very large system sizes (i.e. containing at least on the order of $10^{7}$ spins) 
where finite size effects are negligible. In this manner, we obtain 
$T_{c}$ to within one part in $10^{4}$.  Alternatively, we may examine the Binder cumulant 
$U_{4} = 1 -  [ \langle m \rangle^{4}  ]/3 [ \langle m^{2} \rangle ]^{2}$.

\begin{figure}
\includegraphics[width=.45\textwidth]{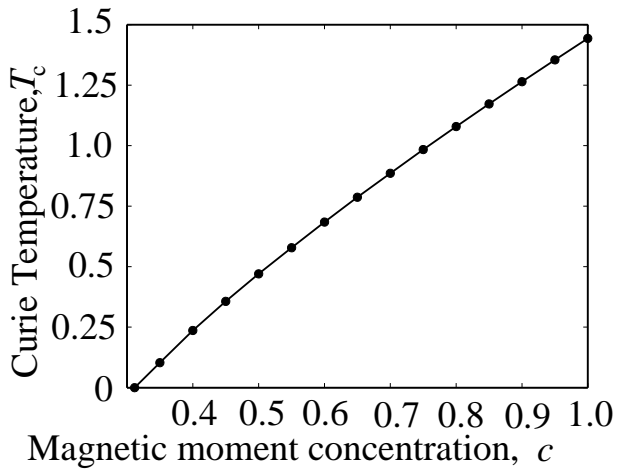}
\caption{\label{fig:Fig1} Curie temperatures calculated from Monte Carlo 
(filled circles) for a range of disorder strengths from the pure case 
$c = 1$ to the site percolation threshold where $c = 0.3116$.}
\end{figure}

Another approach for locating the ferromagnetic transition temperature which we have used and
obtained the 
same Curie temperature results is to examine moderate size systems where finite size effects are 
a more important systematic effect, and to use the Binder cumulant $U_{4}$ in conjunction with the normalized correlation 
length $\xi/L$ to accurately calculate $T_{c}$.
 Finite size
effects preclude a precise determination of $T_{c}$ with either technique alone; the intersections
will actually scale as $T_{c} + A_{U}L^{-1/\nu}$ for the Binder cumulants and 
$T_{c} + A_{\xi} L^{-1/\nu}$ for the normalized correlation length, respectively.  Nevertheless,
by examining two different system size pairs, one may cancel the leading order corrections
from finite size scaling.   In this manner, we have calculate Curie temperatures to within one
part in $10^{4}$ for each impurity concentration we have examined.  $T_{c}$ results are shown in
Fig.~\ref{fig:Fig1} for disorder strengths ranging from the pure cased ($c = 1.0$) to the site percolation threshold
($c = 0.3116$) appropriate to the 3D simple cubic lattice; the Monte Carlo statistical
error is much smaller than the size of the symbols in the graph.
The specific $T_{c}$ values used in the Monte Carlo calculations of singular
thermodynamic quantities appear in Table~\ref{tab:squeeze}; the reciprocals $K_{c} = T_{c}^{-1}$ are given
as well.

\begin{table}
\begin{center}
\begin{tabular}{|c|c|c|}
\hline
concentration  & $T_{c}$  (units of $J_{0}/k_{\mathrm{B}}$) & $K_{c}$ (units of $k_{\mathrm{B}}/J_{0}$)   \\
\hline
$c = 1.0$ & 1.4430  & 0.6930  \\
\hline
$c = 0.95$ & 1.3543 & 0.7384  \\
\hline
$c = 0.90$ & 1.2641 & 0.7911 \\
\hline
$c = 0.80$ & 1.0787  & 0.92705 \\
\hline
$c = 0.70$ & 0.88590 & 1.1288  \\
\hline
$c = 0.60$ & 0.6840 & 1.462 \\
\hline
$c = 0.50$ & 0.4701 & 2.127  \\
\hline
$c = 0.40$ & 0.2361  & 4.235  \\
\hline
\end{tabular}
\caption{\label{tab:squeeze} $T_{c}$ values (with the uncertainty in the last digit for each concentration $c$ given).
The $T_{c}$ results are given in units of $J_{0}/k_{\mathrm{B}}$, whereas the inverse temperatures $K_{c}$ are 
expressed in terms of $k_{\mathrm{B}}/J_{0}$.} 
\end{center}
\vspace{-0.6cm}
\end{table}

The calculation of critical exponents involves the exploitation of finite size scaling trends easily 
obscured by statistical fluctuations stemming from the random character of the disorder, and hence 
it is necessary to average over many realizations of disorder, $10^{5}$ for $c < 0.9$, and at least   
$4 \times 10^{4}$ for weak disorder where $c = 0.9$ and $c = 0.95$, as well as the pure case where 
$c = 1.0$.  The large-scale Monte Carlos calculations have a significant parallel element, and we have
benefited from the use of the HPCC (High Performance Computing Cluster) at the University of Maryland,
cumulatively using approximately a CPU decade to complete the calculations we report on here.

To circumvent critical slowing down plaguing local update techniques such as the Metropolis 
method, our Monte-Carlo calculations employ cluster updates to flip large sets of correlated 
spins.  Specifically, we use alternating Wolff cluster~\cite{Wolff} and Swendsen-Wang sweeps~\cite{Swend}, the latter 
being included because the Swendsen-Wang steps ultimately flip every spin, including isolated 
clusters of moments inaccessible to Wolff cluster moves.  The cluster moves operate by 
flipping groups of thermodynamically correlated spins, and are effective even in the vicinity of 
$T_{c}$ where the diverging correlation length $\xi$ would otherwise be associated with a 
much larger Monte-Carlo autocorrelation time, as certainly would be encountered with the use 
of the Metropolis method.  

To reduce the severity of finite size 
effects, we examine cubic systems of size $L$ with periodic boundary conditions.  We use 1000
hybrid sweeps per disorder realization, and equilibration effects are eliminated by   
discarding the first quarter of the Monte Carlo sweeps.  
Monte Carlo calculations require stochastic input, and we use a Mersenne Twister algorithm
to minimize correlations among random numbers and to ensure the period of the 
sequence far exceeds the number of random numbers used over the span of the Monte Carlo simulations.

Thermal derivatives such as $d \xi/dT$ 
need not be calculated via numerical differentiation; it is more convenient instead to use $d 
\langle g \rangle /dT = ( \langle q E \rangle - \langle q \rangle \langle E \rangle )/(k_{\textrm{B}}T)$
obtained by direct differentiation of $\langle q \rangle = \sum_{\mathrm{conf}} q_{\mathrm{conf}} \exp (-E_{\mathrm{conf}}/
k_{\mathrm{B}} T)/Z$, where the sum is over all possible system configurations, $Z$ is the partition function, 
$E$ is the internal energy, and $q$ is a generic thermodynamic variable such as the magnetization.  

\begin{figure}
\includegraphics[width=.45\textwidth]{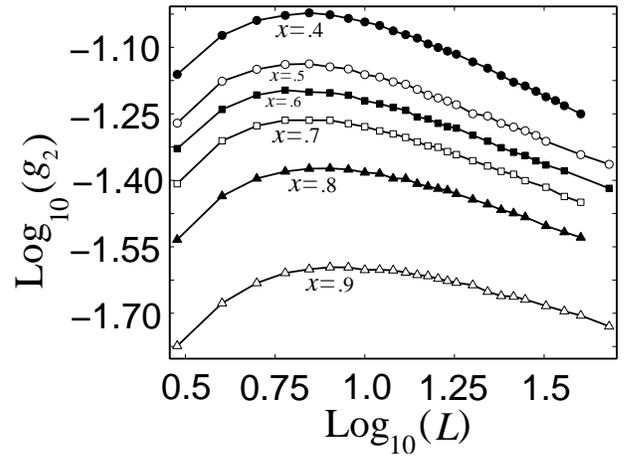}
\caption{\label{fig:Fig2} The graph contains $\log$-$\log$ traces of the self-averaging
parameter $g_{2} = [ \langle m^{2} \rangle^{2} ] - [ \langle m^{2} \rangle ]^{2}$ for the second moment
of the magnetization for very weak ($c = 0.95$) to quite strong disorder ($c = 0.40$); symbols are
from Monte-Carlo calculations, and the solid line is a guide to the eye.}
\end{figure}

By examining the parameter $g_{2}$, which provides a measure of typical fluctuations from one realization of 
disorder to the next, we find clear evidence of self-averaging at the critical temperature $T_{c}$.  Results for 
$g_{2}$ for a range of disorder strengths are shown in Fig.~\ref{fig:Fig2}.
The $\log$-$\log$ $g_{2}$ curves are non-monotonic, increasing for small values of $L$ and attaining a 
maximum (typically for systems containing on the order of 700 spins) before decreasing and  
ultimately becoming linear for sufficiently small system sizes.
An asymptotic power law decay in $L$ of $g_{2}$ for large system sizes is consistent with a monotonic decreases of  
$g_{2}$, a hallmark of self-averaging in the bulk limit.  

A more subtle question is whether disorder has an effect on the critical behavior of the Heisenberg model.
Asymptotic finite size scaling behavior such as $m \propto L^{-\beta/\nu}$,
$\chi \propto \chi_{0} L^{\gamma/\nu}$,  and $d \xi/dT \propto L^{1/\nu}$ imply the
corresponding $\log - \log$ plots will become linear for large enough $L$ with the slope yielding the
critical exponent of interest.  However, 
although singular thermodynamic quantities such as the magnetization $m$ and the susceptibility 
$\chi$ vary asymptotically as $\chi = \chi_{0} L^{\gamma/\nu}$ and $m = m_{0} L^{-\beta/\nu}$, respectively, 
site disorder is a source of important corrections to leading order scaling, which must 
be taken into account 
to obtain accurate expressions for critical exponents such as $\gamma/\nu$ and 
$\beta/\nu$.  Hence, in addition to the amplitude and exponent of the most singular contributions to 
$\chi$ and $m$, we perform nonlinear least squares fitting to take into account the next-to leading order 
exponent and amplitude relative to that of the leading term with 
\begin{align}
\chi (L) = \chi_{0} (L^{\gamma/\nu} + B_{\gamma} L^{\epsilon_{\gamma}});\\
m(L) = m_{0} (L^{-\beta/\nu} + B_{\beta} L^{\epsilon_{\beta}});\\
\frac{d \xi}{dT} (L) = A_{\nu}^{0} (L^{1/\nu} + B_{\nu} L^{\epsilon_{\nu}}),
\end{align} 
where the coefficients $B$ are the relative amplitude of the first correction to primary scaling, and 
the exponents labeled $\epsilon$ are next to leading order exponents. 

We calculate critical exponents and amplitudes by minimizing the sum of the squares of the 
relative differences, e.g. for 
the magnetic susceptibility exponent $\gamma$, with 
$\sigma = \tfrac{1}{N} \left [ \displaystyle{\sum_{i=1}^{N}} \left ( \tfrac{ \gamma_{L_{i}}^{\mathrm{MC}} - 
\gamma_{L_{i}}^{\mathrm{LSF}}}{\gamma_{L_{i}}^{\mathrm{MC}}} 
 \right)^{2} \right]^{1/2}$, where $\gamma_{L_{i}}$ is calculated numerically with Monte 
Carlo simulations and $\gamma_{L_{i}}^{\mathrm{LSF}}$ is given in Eq. 1 for the system size $L_{i}$.  To    
carry out the nonlinear least squares fitting, we use a stochastic algorithm with an annealing stage (i.e. the Metropolis 
Criterion is used with the quantity $\sigma$ treated as an ``energy'' and the ``temperature'' reduced 
at a linear rate in the number of Monte Carlo sweeps over the exponents and amplitudes) 
to minimize $\sigma$ by randomly perturbing exponents and amplitudes;
after the annealing phase, the Monte Carlo moves in the exponent and amplitude space are accepted only 
if the sum of the squares of differences is thereby reduced.  To navigate the shallow ``energy'' landscape
corresponding to $\sigma$, 
the average magnitude of the random shifts is augmented (decreased) by a factor $(1 + \epsilon)$ if a move 
is accepted (rejected) with $\epsilon \sim 10^{-5}$.   In addition, we check for convergence of  
the critical exponents and amplitudes
by successively doubling the time span of the annealing until the results  cease to change.

In experiment, the reduced temperature $t$ is more readily tuned than the system size.  To show 
how the effective critical behavior may vary appreciably for, we calculate the magnetic susceptibility 
$\chi$ for $t > 0$, but in the bulk limit as would be appropriate for comparison experiment. 
For finite $t$, it is sufficient to examine system sizes, such that $L \gg \xi$ since the correlation 
length will be finite for temperatures above $T_{c}$.  We find the condition $\xi/L < 0.06$ is sufficient
to reduce finite size to a negligible level. In addition, by calculating $\chi$ for a number of 
different system sizes, we may correct for finite size effects; we have 
explicitly verified that a 
relation of the form $A + B e^{-\kappa L/xi}$ is a very good approximation to the 
dependence of thermodynamic variable on system size when $L$ is at least on the 
order of a few correlation lengths, a condition we use to further improve our 
approximation to bulk behavior, or to relax somewhat the condition $\xi/L < 0.06$ by examining somewhat 
smaller systems and subsequently removing residual finite size effects.

\begin{figure}
\includegraphics[width=.45\textwidth]{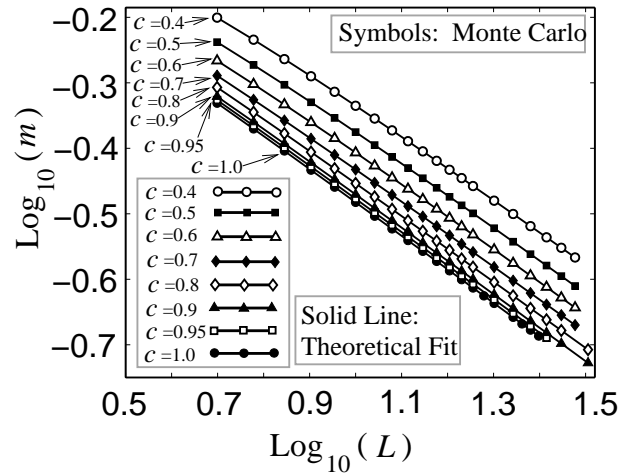}
\caption{\label{fig:Fig3} The image shows $\log$-$\log$ plots of the 
magnetization $m$ versus system size $L$ for the pure Heisenberg model where $c = 1.0$, and 
disorder ranging from quite weak ($c = 0.95$) to very strong 
($c = 0.40$).  The symbols are results from 
Monte Carlo calculations, and the solid lines are theoretical fits.}
\end{figure}

\begin{figure}
\includegraphics[width=.45\textwidth]{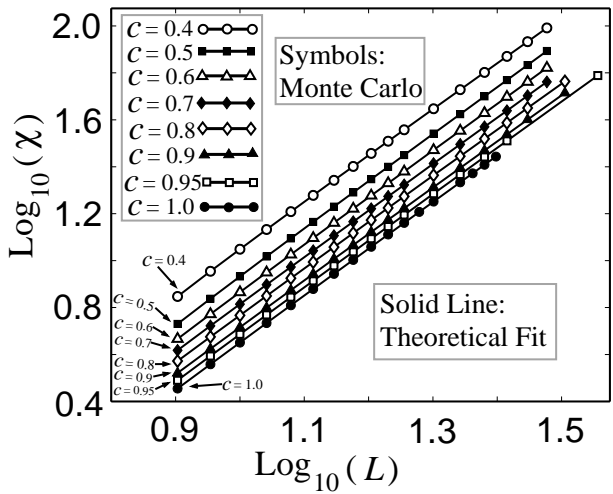}
\caption{\label{fig:Fig4} The graph contains $\log$-$\log$ plots of the 
magnetic susceptibility $\xi$ versus $L$ for the pure Heisenberg model with $c = 1.0$, and 
disorder ranging from quite weak ($c = 0.95$) to very strong ($c = 0.40$).  The symbols are 
results from Monte Carlo calculations, and the solid lines are theoretical fits.}
\end{figure}

\begin{figure}
\includegraphics[width=.45\textwidth]{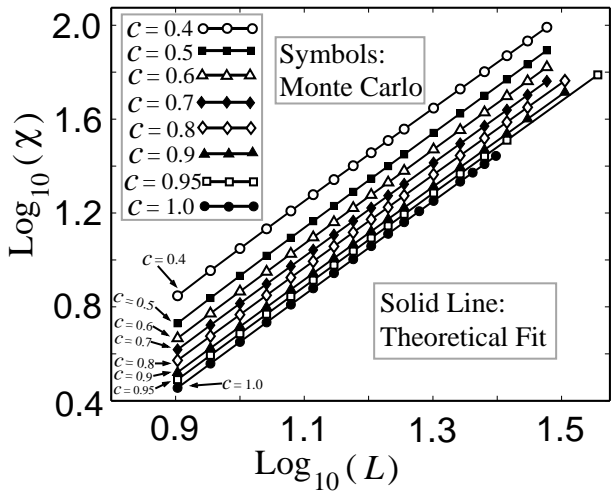}
\caption{\label{fig:Fig5} Shown are $\log$-$\log$ traces of the thermal 
derivative of the correlation length $d \xi/dT$ versus $L$ for the pure Heisenberg model ($c = 1.0$),
and magnetic moment concentrations ranging from very weak ($c = 0.95$) to quite strong disorder ($c = 0.40$). Symbols 
are Monte Carlo data, and the solid lines are theoretical fits.}
\end{figure}

\begin{table}
\begin{center}
\begin{tabular}{|c|c|c|c|c|c|}
\hline
concentration $c$ & $\beta_{\mathrm{pure}}/\nu_{\mathrm{pure}}$ & $\beta/\nu$ & $m_{0}$ & $\epsilon_{\beta}$ & $B_{\beta}$ \\
\hline
$c = 1.0$ & 0.516 & 0.5159 & 1.083 & -2.365 & -0.233\\
\hline
$c = 0.95$ & 0.516 & 0.5150 & 1.096 & -2.242 & -0.2269 \\
\hline
$c = 0.90$ & 0.516 & 0.5143 & 1.114 & -1.975 & -0.2001 \\
\hline 
$c = 0.80$ & 0.516 & 0.5080 & 1.142 & -2.038 & -0.2538 \\
\hline
$c = 0.7$ & 0.516 & 0.5106 & 1.221 & -1.648 & -0.2651 \\
\hline
$c = 0.6$ & 0.516 & 0.5248 & 1.386 & -1.305 & -0.3164 \\
\hline
$c = 0.5$ & 0.516 & 0.5233 & 1.485 & -1.373 & -0.3779 \\
\hline
$c = 0.4$ & 0.516 & 0.5011 & 1.498 & -1.919 & -0.5600 \\
\hline
\end{tabular}
\caption{\label{tab:Tab1} Critical exponent ratios $\beta/\nu$ for the magnetization with the 
amplitude $m_{0}$ of the leading order term, the exponent $\epsilon_{\beta}$ of the first 
correction to primary scaling, and the relative amplitude $B_{\beta}$ of the next-to leading 
order term.} 
\end{center}
\vspace{-0.6cm}
\end{table}
  
\begin{table}
\begin{center}
\begin{tabular}{|c|c|c|c|c|c|}
\hline
concentration $c$ & $\gamma_{\mathrm{pure}}/\nu_{\mathrm{pure}}$ & $\gamma$ & $\chi_{0}$ & $\epsilon_{\gamma}$ & $B_{\gamma}$ \\
\hline
$c = 1.0$ & 1.955 & 1.957 & 0.05206 & 0.9468 & -0.5276 \\
\hline
$c = 0.95$ & 1.955 & 1.963 & 0.05426 & 0.2783 & -1.309 \\
\hline
$c = 0.90$ & 1.955 & 1.954 & 0.0598 & 0.4372 & -1.054 \\
\hline
$c = 0.80$ & 1.955 & 1.935 & 0.0724 & 0.9121 & -0.6482\\
\hline
$c = 0.70$ & 1.955 & 1.973 & 0.07037 & -0.7585 & -7.574 \\
\hline
$c = 0.60$ & 1.955 & 1.977 & 0.08014  & 0.2133 & -1.957 \\
\hline
$c = 0.50$ & 1.955 & 1.993 & 0.08915 & -0.8402 & -15.86 \\
\hline
$c = 0.40$ & 1.955 & 1.945 & 0.1317 & -0.7112 & -15.91 \\
\hline
\end{tabular}
\caption{\label{tab:Tab2} Critical exponent ratios $\gamma/\nu$ for the magnetization with the 
 amplitude $\chi_{0}$ of the leading order term, the exponent $\epsilon_{\gamma}$ of the first 
correction to primary scaling, and the relative amplitude $B_{\gamma}$ of the next-to
leading order term.}
\end{center}
\vspace{-0.6cm}
\end{table}

\begin{table}
\begin{center}
\begin{tabular}{|c|c|c|}
\hline
concentration $c$ & $\eta_{\mathrm{pure}}$ & $\eta$  \\
\hline
$c = 1.0$ & 0.038 & 0.043  \\
\hline
$c = 0.95$ & 0.038 & 0.037  \\
\hline
$c = 0.90$ & 0.038 & 0.046 \\
\hline
$c = 0.80$ & 0.038 & 0.065 \\
\hline
$c = 0.70$ & 0.038 & 0.027  \\
\hline
$c = 0.60$ & 0.038 & 0.023 \\
\hline
$c = 0.50$ & 0.038 & 0.007  \\
\hline
$c = 0.40$ & 0.038 & 0.046  \\
\hline
\end{tabular}
\caption{\label{tab:Tab3} Critical exponents $\eta$ for disorder ranging from the 
pure case ($c = 1.0$) to strong disorder where $c = 0.4$.}
\end{center}
\vspace{-0.6cm}
\end{table}

\begin{table}[h]
\begin{center}
\begin{tabular}{|c|c|c|c|c|c|}
\hline
concentration $c$ & $\nu_{\mathrm{pure}}$ & $\nu$ & $A_{\nu}^{0}$& $\epsilon_{\nu}$ & $r_{\nu}$ \\
\hline
$c = 1.0$ & 0.714 & 0.7149 & 0.2862 & -1.428 & 2.798 \\
\hline
$c = 0.95$ & 0.714 & 0.7291 & 0.2500 & -1.2543 & 1.760 \\
\hline
$c = 0.9$ & 0.714 & 0.7335 & 0.2045 & 0.4941 & 0.2366 \\
\hline
$c = 0.80$ & 0.714 & 0.7412 & 0.1322 & 0.7347  & 0.3706 \\
\hline
$c = 0.70$ & 0.714 & 0.7428 & 0.07814 & 0.7675 & 0.5878 \\
\hline
$c = 0.60$ & 0.714 & 0.7018 & 0.02545 & 0.9572 & 1.923 \\
\hline
$c = 0.50$ & 0.714 & 0.7188 & 0.01330 & 0.8838 & 1.759 \\
\hline
$c = 0.40$ & 0.714 & 0.6997 & 0.00261 & 0.6612 & 4.485 \\
\hline
\end{tabular}
\caption{\label{tab:Tab4} Correlation length critical exponents $\nu$ for $d \xi /dT$ with the amplitude
$A_{\nu}^{0}$ of the leading order term, the exponent $\epsilon_{\nu}$ of the first correction to scaling, and the 
relative amplitude $r_{\nu}$ of the first correction term.}
\end{center}
\vspace{-0.6cm}
\end{table}

\section{Critical Behavior of the Disordered Heisenberg Model}

   The $\log$-$\log$ plots in Fig.~\ref{fig:Fig3} show the magnetization with symbols representing 
Monte Carlo results, and the continuous curves are obtained from the corresponding nonlinear 
least squares fits.  The excellent agreement of the Monte Carlo data and theoretical fits may 
also be seen in the Appendix, where the simulation data and theoretical results are given 
to five significant figures.  Similarly, the magnetic susceptibilities appear in Fig.~\ref{fig:Fig4}, where 
symbols represent the Monte Carlo results and solid lines obtained from theoretical fits closely match the Monte Carlo
data.  Finally, the correlation length thermal derivatives $d \xi/dT$ are graphed in Fig.~\ref{fig:Fig5}, and there is 
again good agreement between Monte Carlo results (symbols) and the solid lines obtained from theoretical results.

Exponents and critical amplitudes are given for $\beta/\nu$ in Table~\ref{tab:Tab1}, $\gamma/\nu$ (corresponding
to the susceptibility) in Table~\ref{tab:Tab2}, $\eta$ in Table~\ref{tab:Tab3}, and $\nu$ in Table~\ref{tab:Tab4}. 
The exponent $\eta$ is calculated from $\gamma$ and $\nu$ with $\eta = 2  \gamma/\nu$.  The parameter $\eta$
is a sensitive parameter and, accordingly, there is greater variance in the results.  However, the $\eta$ 
values listed in Table~\ref{tab:Tab3} each have the same positive sign irrespective of the strength of the 
site disorder. The leading 
order exponents are consistent with those of the pure Heisenberg Universality class with deviations 
due only to statistical Monte Carlo error, not systematic effects 
related to the disorder strength.  
Hence, since each of the exponents $\beta$, $\nu$, $\gamma$, and $\eta$ are stable with respect 
to the introduction of site defects, we conclude for the Heisenberg model that critical 
behavior is unchanged even in the presence of very strong disorder. 

\section{Effective Critical Behavior and Apparent Violation of the Harris Criterion}

\begin{figure}
\includegraphics[width=.49\textwidth]{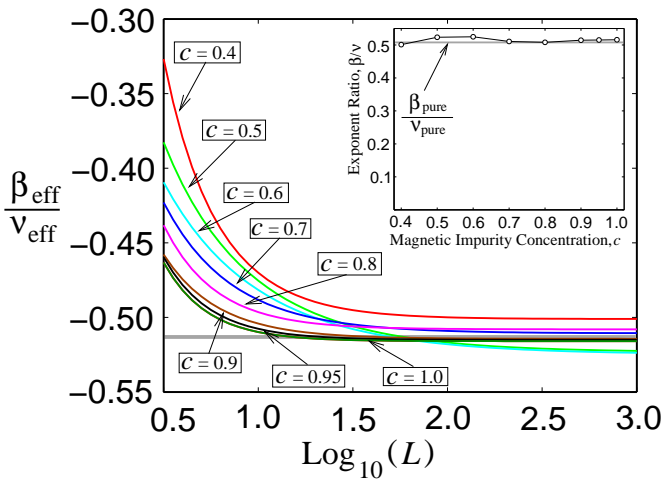}
\caption{\label{fig:Fig6} (color online) Effective $\beta/\nu$ curves are shown in a semi-logarithmic 
graph for various impurity concentrations $c$ 
for several decades of $L$.  The inset shows the calculated $\beta/\nu$ values versus the
impurity concentration.  The horizontal gray lines in both the primary graph and the 
inset correspond to $\beta/\nu$ for the pure system.}

\end{figure}

\begin{figure}
\includegraphics[width=.49\textwidth]{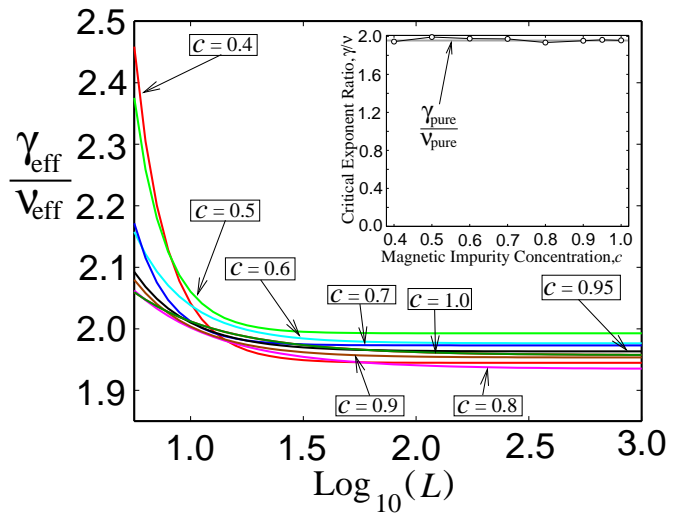}
\caption{\label{fig:Fig7} (color online) Effective $\gamma/\nu$ curves are graphed in a semi-logarithmic 
plot for various impurity concentrations $c$ for several decades of system sizes, with 
the inset showing $\gamma/\nu$ versus the impurity concentration $c$. The horizontal gray lines 
in the main figure and graph inset indicate $\gamma_{\mathrm{pure}}/\nu_{\mathrm{pure}}$ for the case
where there is no disorder.}
\end{figure}

\begin{figure}
\includegraphics[width=.49\textwidth]{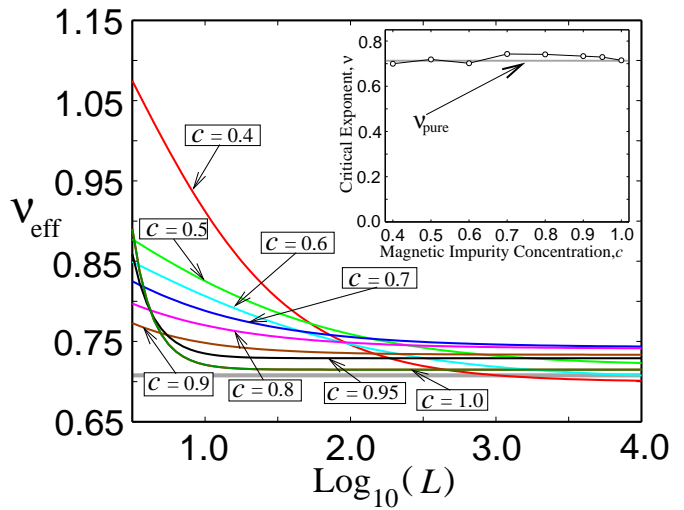}
\caption{\label{fig:Fig8} (color online) Effective $\nu$ curves are plotted in a semi-logarithmic fashion for 
a range of magnetic moment concentrations $c$ for a few decades of $L$; the inset is a graph of 
the calculated exponent $\nu$ versus $c$.  Again, the horizontal gray lines correspond to $\nu_{\mathrm{pure}}$
 for the case $c = 1$.} 
\end{figure}

Although ultimately we find that the critical behavior of the pure 
Heisenberg model emerges as the dominant part of the singular components of thermodynamic 
variables such as the magnetization $m$ and magnetic susceptibility $\chi$, 
finite size effects may obscure the genuine critical behavior for systems of small to moderate size
where bulk critical behavior has not taken hold.
Fig.~\ref{fig:Fig6}, Fig.~\ref{fig:Fig7}, and Fig.~\ref{fig:Fig8} 
show the apparent critical indices 
which would be obtained as the slope $d \log (\chi)/d \log (t) = \tfrac{t}{\chi} \tfrac{d \chi}{dt}$ of 
the log-log graph, a quantity which may differ significantly for the first several decades of the system 
size $L$ before eventually converging to the critical indices of the pure Heisenberg model,     
indicated with horizontal gray lines.  Qualitatively similar behavior has been seen in 
renormalization group (RG) calculations~\cite{Dudka} as well as in experiment~\cite{Fahnle,Kaul1,Kaul2,Kaul3,Babu,Perumal}
 with the reduced 
temperature $t$ varied instead of the system size $L$. The insets of Fig.~\ref{fig:Fig6}, Fig.~\ref{fig:Fig7}, 
and Fig.~\ref{fig:Fig8} display $\beta/\nu$,
$\gamma/\nu$, and $\nu$ obtained from the nonlinear least squares fits.  Again, throughout the broad 
disorder spectrum considered, even for very strongly 
strongly disordered systems (e.g. for the case $c = 0.4$), the critical indices we 
calculate are compatible with those of the 
pure system where there is no disorder.

\begin{figure}
\centerline{\includegraphics[width=3.2in]{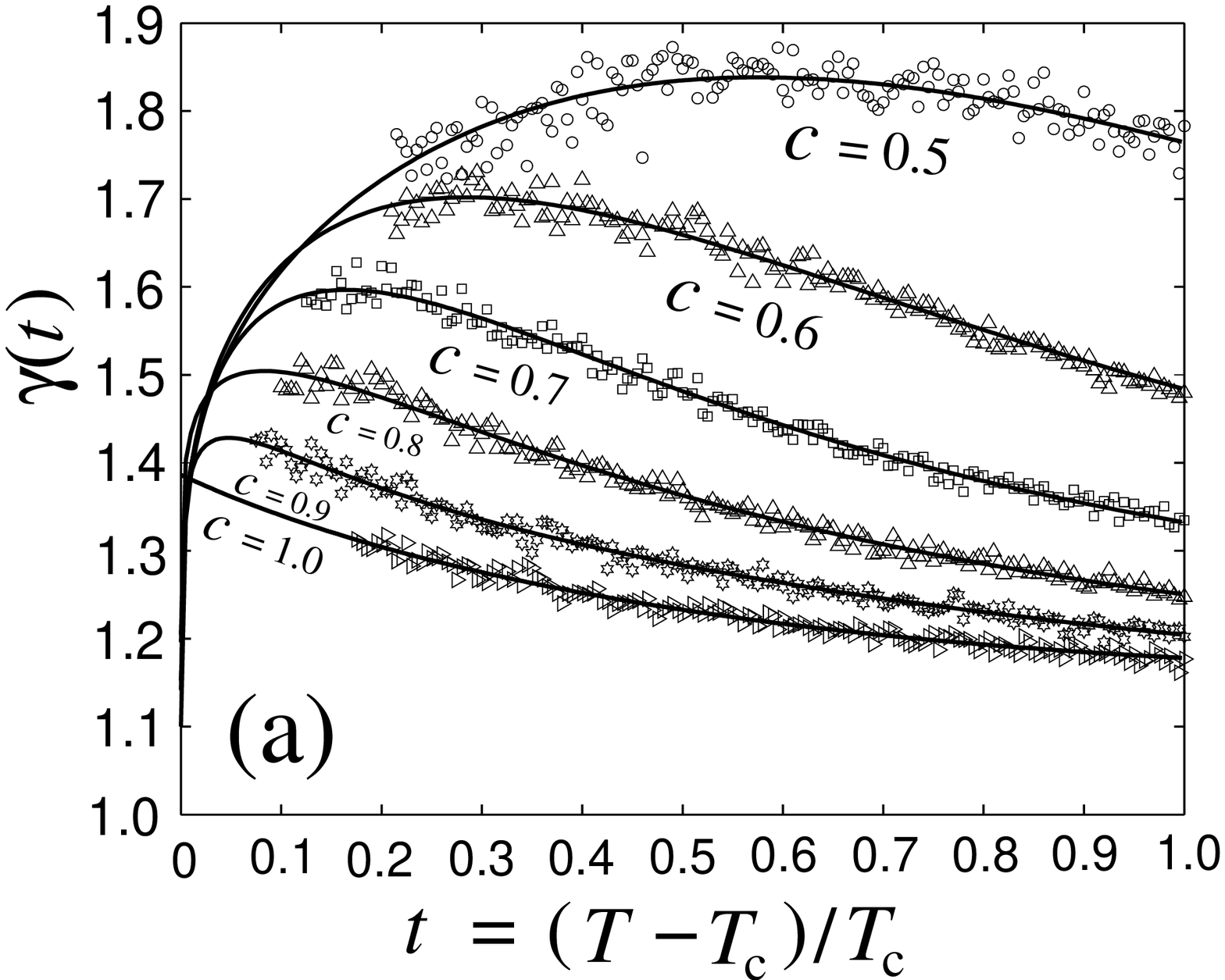}}
\vspace{.5cm}
\centerline{\includegraphics[width=2.8in]{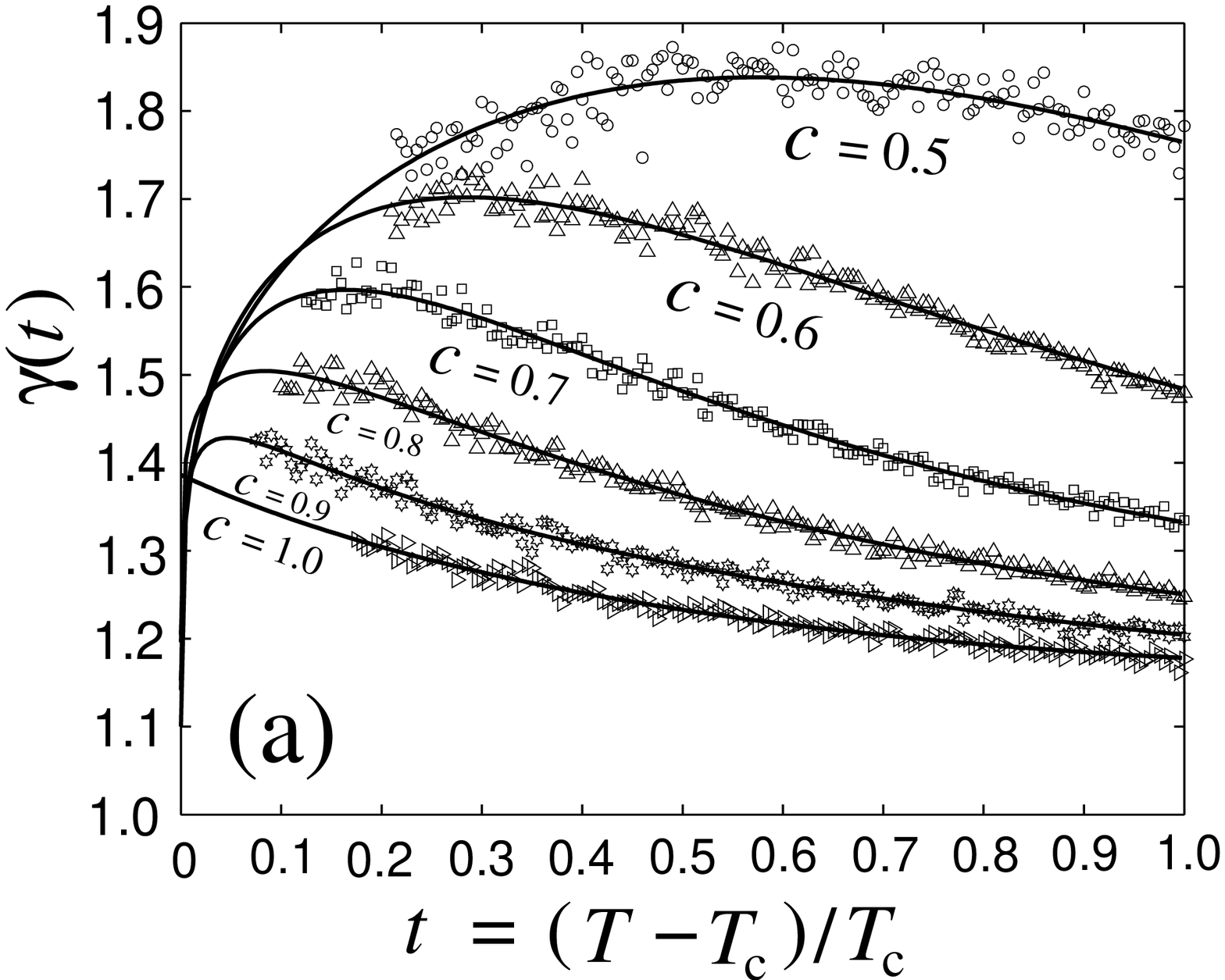}}
\caption{Panel (a) shows the effective magnetic susceptibility exponent $\gamma(t)$ versus the
reduced temperature $t$.  The solid curves are analytical results from the least square
fitting, and the symbols represent Monte Carlo data.  In panel (b), solid curves
are again obtained from least squares fitting, and the symbols are Monte Carlo
data.}
\label{fig:Fig9}
\end{figure}

To make direct contact with experiment and show explicitly the apparent change in critical behavior may be set up by 
strong disorder, we show finite $t$ results where the effective critical exponent $\gamma_{\mathrm{eff}}$  
corresponding to the magnetic susceptibility has been calculated; results are shown in panel (a) and 
panel (b) of Fig.~\ref{fig:Fig9}.  For $t > 0$, the susceptibility $\chi$ will scale as 
$\chi = \chi_{0} ( t^{-\gamma} + B t^{y_{1}} + C t^{y_{2}} + D t^{y_{3}} + \ldots)$ where 
$\gamma$ is the genuine critical exponent for $\chi$, and the terms with exponents such as $y_{1}$, 
$y_{2}$ for the first two subleading terms are corrections to scaling which may have a significant 
effect if $t$ is sufficiently large or in the presence of strong enough disorder.

In the critical regime where subleading terms may be neglected, one may compute, e.g. for $\chi$, 
$\gamma = -\tfrac{t}{\chi} d \chi/dt$.  However, further from $T_{c}$ where corrections to singular 
critical behavior are more important, one obtains an ``effective'' exponent $\gamma(t)$ given by 
\begin{align}
\gamma (t) \equiv \frac{-t}{\chi}\frac{d \chi}{dt} = \gamma \left( \frac{1 + By_{1}t^{y_{1} +
\gamma} + Cy_{2}t^{y_{2} + \gamma} + \ldots}{1 + Bt^{y_{1} + \gamma} + Ct^{y_{2} + \gamma} + \ldots}\right);
\label{Eq:eq101}
\end{align}
$\gamma(t)$ will eventually tend to the leading order exponent $\gamma$ as $t \rightarrow 0$, though one may have to 
measure $\chi$ at very low values of $t$ if there is a strong disorder component.

The graphs shown in Fig.~\ref{fig:Fig9} show results from two distinct calculations of $\gamma(t)$.      
In panel (a) of Fig.~\ref{fig:Fig9}, the Monte Carlo data is drawn from a study where fewer disorder 
realizations are examined (though still at least $5 \times 10^{3}$ configurations of disorder
are analyzed)
in favor of obtaining a larger data set; Monte Carlo results are 
shown as symbols with theoretical curves obtained from nonlinear least square fitting shown on the 
same graph.  Similarly, for the set of calculations involving  
fewer data points but more intensive disorder averaging, Monte Carlo data is graphed 
as symbols in panel (b) of Fig.~\ref{fig:Fig9}, while again solid lines are theoretical curves gleaned from 
least squares fitting.

In both cases, although $\gamma(t)$ for the pure $(c = 1)$ case rises steadily with decreasing 
$t$, the curves for each of the disordered systems are nonmonotonic; the initial rise with 
decreasing $t$ is followed by a peak and subsequent decline to the asymptotic value of 
$\gamma$ only for very small values of the reduced temperature on the order of 
$t \sim 10^{-3}$.

We reiterate that the critical exponents we calculate are consistent with 
$\nu > 2/3$ where disorder is irrelevant to the universality class in the Renormalization Group (RG) sense
This inequality has been placed on a rigorous footing in 
theoretical work~\cite{Chayes} under a broad range of conditions, 
and has also been established for correlated
disorder~\cite{Weinrib}.  We also emphasize that while the genuine Heisenberg model
critical exponents satisfy the hyperscaling relations, the apparent critical
exponents obtained away from the critical behavior are not consistent with 
the hyperscaling formulas, an indication of their problematic nature.

\section{Conclusions}

In conclusion, within a large-scale Monte Carlo study, we have examined Heisenberg models on three dimensional lattices with 
randomly deleted magnetic moments as a source of disorder, finding self-averaging to be intact as predicted by the Harris      
criterion.  Moreover, our finite size scaling studies show leading order critical behavior not to be 
influenced by the presence of random defects, 
with critical exponents identical to those of the pure Heisenberg model universality class even 
for very strong disorder in the vicinity of the site percolation threshold where long-range ferromagnetic order is lost 
altogether for $T < 0$.  However, while the leading order exponents are not sensitive 
to disorder, the presence of site defects sets up corrections to primary scaling which 
skew the effective exponents for finite system sizes $L$, a characteristic which might na\"{i}vely be regarded as 
evidence for the violation of the Harris criterion.  A qualitatively similar apparent violation of the Harris criterion is seen in 
experiment where thermodynamic quantities such as the magnetic susceptibility are measured with respect to the 
reduced temperature $t$, and we have also calculate the same quantities in the bulk limit for $t > 0$, finding the 
same apparent violation of the Harris criterion. 
We conclude by asserting the asymptotic validity of the Harris criterion sufficiently close to the critical 
temperature in the strongly disordered Heisenberg model appropriate for diluted magnetic semiconductors, at 
the same time pointing out that slightly away from the critical temperature, the 
effective exponents may very well reflect an apparent (and incorrect) violation unless extremely careful 
measures are taken to include finite-size scaling and corrections to scaling in the analyses

\section{Appendix: Thermodynamic Quantities from Monte Carlo and Analytical Fits}
The appendix contains a sequence of tables explicitly giving thermodynamic quantities calculated 
in Monte Carlo simulations with the theoretical fits obtained by stochastically 
enhanced least squares fitting.  The theoretical results are in very close agreement with the 
corresponding Monte Carlo data. 
\begin{table}
\begin{center}
\begin{tabular}{|c|c|c|c|c|c|c|}
\hline
$n$ & $m^{x=1.0}_{\mathrm{num}}$ & $m^{x=1.0}_{\mathrm{fit}}$ & 
$\chi^{x=1.0}_{\mathrm{num}}$ & $\chi^{x=1.0}_{\mathrm{fit}}$ & 
$\frac{d\xi}{dT}^{x = 1.0}_{\mathrm{num}}$ & $\frac{d\xi}{dT}^{x = 1.0}_{\mathrm{fit}} $   \\
\hline
$5$ & 0.46661 & 0.46662 & & & & \\
\hline
$6$ & 0.42624 & 0.42620 & & & & \\
\hline
$7$ & 0.39444 & 0.39445 & & & & \\
\hline
$8$ & 0.36869 & 0.36871 &2.8510 &2.8512 & 5.2874 & 5.2882 \\
\hline
$9$ & 0.34730 & 0.34731 &3.6186 & 3.6181 & 6.2244 & 6.2216 \\
\hline
$10$ & 0.32915 & 0.32917 &4.4759 & 4.4740 & 7.2043 & 7.1992 \\
\hline
$11$ & 0.31357 & 0.31355 & 5.4184   & 5.4148   & 8.2064 & 8.2178 \\
\hline
$12$ & 0.29991 & 0.29991 & 6.4509 & 6.4481 & 9.2678 & 9.2749 \\
\hline
$13$ & 0.28786 & 0.28787 & 7.5750 & 7.5712 & 10.372 & 10.369 \\
\hline 
$14$ & 0.27717 & 0.27714 & 8.7794 & 8.7790 & 11.495 & 11.497 \\
\hline   
$15$ & 0.26754 & 0.26751  & 10.075 & 10.074 & 12.677 & 12.658 \\
\hline
$16$ & 0.25880  & 0.25879  & 11.460 & 11.456 & 13.853 & 13.851 \\
\hline
$17$ &  0.25084 & 0.25086  & 12.924 & 12.924 & 15.065 & 15.074  \\
\hline
$18$ & 0.24358 & 0.24360 & 14.480 & 14.480 & 16.331 & 16.326 \\
\hline
$19$ & 0.23690 & 0.23693 & 16.120 & 16.121 & 17.618 & 17.607 \\
\hline
$20$ & 0.23077 & 0.23076 & 17.844 & 17.848 & 18.906 & 18.915 \\
\hline
$22$ & 0.21972 & 0.21972 & 21.547 & 21.560 & 21.586 & 21.609 \\
\hline
$23$ & 0.214763 & 0.21475 & 23.559 & 23.545 & 22.996 & 22.994 \\
\hline
$24$ & 0.21013 & 0.21010 & 25.609 & 25.614 & 24.391 & 24.403 \\
\hline
$25$ & 0.20573  & 0.20573 & 27.774 & 27.769 & 25.859 & 25.836 \\
\hline 
\end{tabular}
\caption{\label{tab:Tab5}Thermodynamic quantities from Monte Carlo simulations and 
theoretical fits for the pure system, where $c = 1.0$.} 
\end{center}
\vspace{-0.6cm}
\end{table}

\begin{table}
\begin{center}
\begin{tabular}{|c|c|c|c|c|c|c|}
\hline
$n$ & $m^{x=0.95}_{\mathrm{num}}$ & $m^{x=0.95}_{\mathrm{fit}}$ &
$\chi^{x=0.95}_{\mathrm{num}}$ & $\chi^{x=0.95}_{\mathrm{fit}}$ &
$\frac{d\xi}{dT}^{x = 0.95}_{\mathrm{num}}$ & $\frac{d\xi}{dT}^{x = 0.95}_{\mathrm{fit}} $   \\
\hline
$5$ & 0.47156 & 0.47162 & & & & \\
\hline
$6$ & 0.43115 & 0.43101 & & & & \\
\hline
$7$ & 0.39900 & 0.39908 & & & & \\
\hline
$8$ & -.37323 & 0.37317 & 3.0920 & 3.0911 & 4.3618 & 4.3626 \\
\hline 
$9$ & 0.35155 & 0.35161 & 3.9221 & 3.9241 & 5.1192 & 5.1173 \\
\hline
$10$ & 0.33341 & 0.33332  & 4.8492 & 4.8521 & 5.9058 & 5.9050 \\
\hline
$11$ & 0.31749 & 0.31756 & 5.8821 & 5.8747 & 6.7220 & 6.7234 \\
\hline 
$12$ & 0.30387 & 0.30380 & 6.9917 & 6.9915 & 7.5733 & 7.5706 \\
\hline
$13$ & 0.29147 & 0.29165 & 8.2009 & 8.2022 & 8.4337 & 8.4448 \\
\hline
$14$ & 0.28075 & 0.28082 & 9.4934 & 9.5065 & & \\
\hline
$15$ & 0.27108 & 0.27109 & 10.912 & 10.904 & 10.278 & 10.269 \\
\hline
$16$ & 0.26227 & 0.26228 & 12.396 & 12.395 & 11.216 & 11.217 \\
\hline
$17$ & 0.25435 & 0.25427 & 13.983  & 13.978 & 12.173 & 12.188 \\
\hline
$18$ & 0.24699 & 0.24693 & 15.666  & 15.655 & 13.197 & 13.179 \\
\hline
$20$ & 0.23401 & 0.23395 & 19.286 & 19.284 & 15.242 & 15.225 \\
\hline
$22$ & 0.22290 & 0.22279 & 23.237 & 23.281 & 17.323 & 17.349 \\
\hline
$24$ & 0.21295 & 0.21306 & 27.658 & 27.646 & 19.546 & 19.545 \\
\hline
$26$ & 0.20443 & 0.20448 & 32.391 & 32.376 & 21.816 & 21.811 \\
\hline
$36$ & & & 61.475 & 61.475 & & \\
\hline
\end{tabular}
\caption{\label{tab:Tab6} Monte-Carlo results and theoretical fits for 
thermodynamic quantities for a weakly disordered Heisenberg model with 
$c = 0.95$.}
\end{center}
\vspace{-0.6cm}
\end{table}

\begin{table}
\begin{center}
\begin{tabular}{|c|c|c|c|c|c|c|}
\hline
$n$ & $m^{x=0.9}_{\mathrm{num}}$ & $m^{x=0.9}_{\mathrm{fit}}$ &
$\chi^{x=0.9}_{\mathrm{num}}$ & $\chi^{x=0.9}_{\mathrm{fit}}$ &
$\frac{d\xi}{dT}^{x = 0.9}_{\mathrm{num}}$ & $\frac{d\xi}{dT}^{x = 0.9}_{\mathrm{fit}} $   \\
\hline
$5$ & 0.47733 & 0.47747 & & & & \\
\hline
$6$ & 0.43701 & 0.43671  & & & & \\
\hline
$7$ & 0.40455 & 0.40463 & & & & \\
\hline
$8$ & 0.37850 & 0.37856 & 3.3201 & 3.3184 & 2.4417 & 2.4127 \\
\hline
$9$ & 0.35689 & 0.35685 & 4.2089 & 4.2092 & 2.8128 & 2.8098 \\
\hline
$10$ & 0.33837  & 0.33842  & 5.1921 & 5.2010 & 3.2218 & 3.2212 \\
\hline
$11$ & 0.32250 & 0.32253 & 6.2947 & 6.2933 & 3.6437 & 3.6461 \\
\hline
$12$ & 0.30863 & 0.30863 & 7.4897 & 7.4857 & 4.0835 & 4.0835 \\
\hline
$13$ & 0.29638 & 0.29636 & 8.7799 & 8.7776 & 4.5272 & 4.5330 \\
\hline
$14$ & 0.28531 & 0.28541 & 10.1715 & 10.1687 & 4.9940 & 4.9938 \\
\hline
$15$ & 0.27551 & 0.27558 & 11.670 & 11.659 & 5.4648 & 5.4655 \\
\hline
$16$ & 0.26670 & 0.26667 & 13.245 & 13.247 & 5.9519 & 5.9475 \\
\hline
$17$ & 0.25862 & 0.25856 & 14.928 & 14.933 & 6.4421 & 6.4396 \\
\hline
$18$ & 0.25126 & 0.25113 & 16.714 & 16.718 & 6.9451 & 6.9412 \\
\hline
$20$ & 0.23784 & 0.23799 & 20.604 & 20.579 & 7.9753 & 7.9719 \\
\hline
$22$ & 0.22676 & 0.22668 & 24.797 & 24.828 & 9.0375 & 9.0373 \\
\hline
$24$ & 0.21689 & 0.21681 & 29.428 & 29.464 & 10.136 & 10.135  \\
\hline
$26$ & 0.20809 & 0.20811 & 34.484 & 34.484 & 11.253 & 11.264 \\
\hline 
$28$ & 0.20038 & 0.20036 & 39.888 & 39.906 & 12.411 & 12.422 \\
\hline 
$32$ & 0.18796 & 0.18712 & 51.840 & 51.863 & 14.835 & 14.821 \\
\hline 
\end{tabular}
\caption{\label{tab:Tab7}Monte Carlo data and corresponding theoretical fits for 
the magnetization, magnetic susceptibility, and 
$d \xi/dT$ for mildly disordered Heisenberg models ($c = 0.9$).}   
\end{center}
\vspace{-0.6cm}
\end{table}

\begin{table}
\begin{center}
\begin{tabular}{|c|c|c|c|c|c|c|}
\hline
$n$ & $m^{x=0.8}_{\mathrm{num}}$ & $m^{x=0.8}_{\mathrm{fit}}$ &
$\chi^{x=0.8}_{\mathrm{num}}$ & $\chi^{x=0.8}_{\mathrm{fit}}$ &
$\frac{d\xi}{dT}^{x = 0.8}_{\mathrm{num}}$ & $\frac{d\xi}{dT}^{x = 0.8}_{\mathrm{fit}} $   \\
\hline
$5$ & 0.49310 & 0.49316 & &  & & \\
\hline
$6$ & 0.45209 & 0.45195 & & & & \\
\hline
$7$ & 0.41946 & 0.41937 & & & & \\
\hline
$8$ & 0.39240 & 0.39282 & 3.7346 & 3.7337 & 2.4117 & 2.4127 \\
\hline
$9$ & 0.37079 & 0.37065 & 4.7334 & 4.7340 & 2.8128 & 2.8098 \\
\hline
$10$ & 0.35190 & 0.35180 & 5.8429 & 5.8480 & 3.2218 & 3.2212 \\
\hline
$11$ & 0.33547 & 0.33552 & 7.0858 & 7.0751 & 3.6437 & 3.6461 \\
\hline
$12$ & 0.32145 & 0.32127 & 8.4009 & 8.4145 & 4.0835 & 4.0835 \\
\hline
$13$ & 0.30867 & 0.30867 & 9.8800 & 9.8655 & 4.5272 & 4.5330 \\
\hline
$14$ & 0.29742 & 0.29743 & 11.417 & 11.427 & 4.9940 & 4.9938 \\
\hline
$15$ & 0.28726 & 0.28731 & 13.090 & 13.100 & 6.4421 & 6.4396 \\
\hline
$16$ & 0.27804 & 0.27815 & 14.888 & 14.882 & 5.9519 & 5.9475 \\
\hline
$17$ & 0.26980 & 0.26981 & 16.799 & 16.774 & 6.4421 & 6.4396 \\
\hline
$18$ & 0.26212 & 0.26216 & 18.776 & 18.775 & 6.9451 & 6.9412 \\
\hline
$20$ & 0.24868 & 0.24861 & 23.101 & 23.103 & 7.9753 & 7.9719 \\
\hline
$22$ & 0.23693 & 0.23694 & 27.865 & 27.862 & 9.0375 & 9.0373 \\
\hline
$24$ & 0.22681 & 0.22676 & 33.003 & 33.050 & 10.136 & 10.135 \\
\hline
$26$ & 0.21781 & 0.21777 & 38.657 & 38.664 & 11.253 & 11.264 \\
\hline 
$28$ & 0.20973 & 0.20977 & 44.733 & 44.702 & 12.411 & 12.422  \\
\hline
$32$ & 0.19605 & 0.19607 & 58.050 & 58.043 & 14.835 & 14.821\\
\hline
\end{tabular}
\caption{\label{tab:Tab8}Monte Carlo results with theoretical fits
for key thermodynamic quantities in the case of moderate disorder, where $c = 0.8$.}
\end{center} 
\vspace{-0.6cm}
\end{table}
  
\begin{table}
\begin{center}
\begin{tabular}{|c|c|c|c|c|c|c|}
\hline
$n$ & $m^{x=0.7}_{\mathrm{num}}$ & $m^{x=0.7}_{\mathrm{fit}}$ &
$\chi^{x=0.7}_{\mathrm{num}}$ & $\chi^{x=0.7}_{\mathrm{fit}}$ &
$\frac{d\xi}{dT}^{x = 0.7}_{\mathrm{num}}$ & $\frac{d\xi}{dT}^{x = 0.7}_{\mathrm{fit}} $   \\
\hline 
$5$ & 0.51371 & 0.51391 & & & & \\
\hline
$6$ & 0.47241 & 0.47211 & & & & \\
\hline
$7$ & 0.4389 & 0.43889 & & & & \\
\hline
$8$ & 0.41201 & 0.41168 & 4.1464 & 4.1492 & 1.5110 & 1.5109 \\
\hline
$9$ & 0.38852 & 0.38890 & 5.2738 & 5.2730 & 1.7535  & 1.7530  \\
\hline
$10$ & 0.36941 & 0.36945 & 6.5313 & 6.5225 & 2.0032  & 2.0033 \\
\hline
$11$ & 0.35261 & 0.35261 & 7.8917 & 7.8978 & 2.2611  & 2.2611  \\
\hline
$12$ & 0.33768 & 0.33785 & 9.4050 & 9.3988 & 2.5235 & 2.5262 \\
\hline
$13$ & 0.32469 & 0.32477 & 11.033 & 11.026 & 2.7974 & 2.7980 \\
\hline
$14$ & 0.31308 & 0.31308 & 12.779 & 12.778 & 3.0780 & 3.0763 \\
\hline
$15$ & 0.30259 & 0.30254 & 14.642 & 14.655 & 3.3598 & 3.3608 \\
\hline
$16$ & 0.29296 & 0.29299 & 16.661 & 16.658 & 3.6553 & 3.6512 \\
\hline
$17$ & 0.28451 & 0.28428 & 18.756 & 18.786 & 3.9476 & 9.9472 \\
\hline
$18$ & 0.27625 & 0.27629 & 21.043 & 21.039 & 4.2492 & 4.2487 \\
\hline
$20$ & 0.26210 & 0.26211 & 25.936 & 25.919 & 4.8674 & 4.8675 \\
\hline
$22$ & 0.24997 & 0.24989 & 31.328 & 31.298 & 5.5083 & 5.5059  \\
\hline
$24$ & 0.23933 & 0.23921 & 37.109 & 37.173 & 6.1555 & 6.1630 \\
\hline
$26$ & 0.22973 & 0.22977 & 43.532 & 43.544 & 6.8363 & 6.8376 \\
\hline
$28$ & 0.22125 & 0.22135 & 50.435 & 50.410 & 7.5259 & 7.5290 \\
\hline
$30$ & 0.21377 & 0.21379 & 57.808 & 57.770 & 8.2433 & 8.2364\\
\hline
\end{tabular}
\caption{\label{tab:Tab9} Monte Carlo results and theoretical fits for 
thermodynamic quantities corresponding to critical exponents 
$\beta$, $\gamma$, and $\nu$ for a range of system sizes $L$ 
for moderate disorder $c = 0.7$.}  
\end{center}
\vspace{-0.6cm}
\end{table}

\begin{table}
\begin{center}
\begin{tabular}{|c|c|c|c|c|c|c|}
\hline
$n$ & $m^{x=0.6}_{\mathrm{num}}$ & $m^{x=0.6}_{\mathrm{fit}}$ &
$\chi^{x=0.6}_{\mathrm{num}}$ & $\chi^{x=0.6}_{\mathrm{fit}}$ &
$\frac{d\xi}{dT}^{x = 0.6}_{\mathrm{num}}$ & $\frac{d\xi}{dT}^{x = 0.6}_{\mathrm{fit}} $   \\
\hline
$5$ & 0.54158 & 0.54165 & & & & \\
\hline
$6$ & 0.49865 & 0.49869 & & & & \\
\hline
$7$ & 0.464646 & 0.464353 & & & & \\
\hline
$8$ & 0.43616 & 0.43612 & 4.6385 & 4.6408 & 0.85119 & 0.85087 \\
\hline
$9$ & 0.41211 & 0.41237 & 5.9258 & 5.9152 & 0.98288 & 98364 \\
\hline
$10$ & 0.39194 & 0.39205 & 7.3248 & 7.3370 & 1.1209 & 1.1206 \\
\hline
$11$ & 0.37460 & 0.37440 & 8.9036 & 8.9058 & 1.2616 & 1.2614 \\
\hline
$12$ & 0.35895 & 0.35889 & 10.622 & 10.621 & 1.4060 & 1.4060 \\
\hline
$13$ & 0.34494 & 0.34512 & 12.513 & 12.483 & 1.5528 & 1.5541 \\
\hline
$14$ & 0.33278 & 33279 & 14.457 & 14.491 & 1.7085 & 1.7056 \\
\hline
$15$ & 0.32167 & 0.32166 & 16.652 & 16.644 & 1.8594 & 1.8604 \\
\hline
$16$ & 0.31144 & 0.31156 & 18.917 & 18.942 & 2.0172 & 2.0183 \\
\hline
$17$ & 0.30233 & 0.30240 & 21.374 & 21.386 & 2.1776 & 2.1792 \\
\hline
$18$ & 0.293935 & 0.293852 & 24.026 & 23.975 & 2.3468 & 2.3430 \\
\hline
$20$ & 0.278766 & 0.27880 & 29.614 & 29.586 & 2.6786 & 2.6790 \\
\hline
$22$ & 0.26589 & 0.26579 & 35.780 & 35.775 & 3.0249 & 3.0257 \\
\hline
$24$ & 0.25447 & 0.25442 & 42.481 & 421.540 & 3.3835 & 3.3825 \\
\hline
$26$ & 0.24429 & 0.24435 & 49.879 & 49.833 & 3.7460 & 3.7490 \\
\hline
$28$ & 0.23538 & 0.23537 & 57.848 & 57.792 & 4.1222 & 4.1246 \\
\hline
$30$ & 0.22724 & 0.22729 & 66.285 & 66.278 & 4.5131 & 4.5091 \\
\hline
\end{tabular}
\caption{\label{tab:Tab10}Monte Carlo and corresponding theoretical fits
for $m$, $\xi$, and $d \xi/dT$ for moderate disorder with $c = 0.6$.}
\end{center}
\vspace{-0.6cm}
\end{table}

\begin{table}
\begin{center}
\begin{tabular}{|c|c|c|c|c|c|c|}
\hline
$n$ & $m^{x=0.5}_{\mathrm{num}}$ & $m^{x=0.5}_{\mathrm{fit}}$ &
$\chi^{x=0.5}_{\mathrm{num}}$ & $\chi^{x=0.5}_{\mathrm{fit}}$ &
$\frac{d\xi}{dT}^{x = 0.5}_{\mathrm{num}}$ & $\frac{d\xi}{dT}^{x = 0.5}_{\mathrm{fit}} $   \\
\hline
$5$ & 0.57792 & 0.57819 & & & & \\
\hline
$6$ & 0.53377 & 0.53361 & & & & \\
\hline
$7$ & 0.49814 & 0.49769 & & & & \\
\hline
$8$ & 0.46811 & 0.46798 & 5.3682 & 5.3740 & 0.38645 & 0.38685 \\
\hline
$9$ & 0.44278 & 0.44290 & 6.8930 & 6.8840 & 0.44592 & 0.44569 \\
\hline
$10$ & 0.42123 & 0.42136 & 8.5446 & 8.5634 & 0.50648 & 0.50621 \\
\hline
$11$ & 0.40260 & 0.40262 & 10.433 & 10.413 & 0.56934 & 0.56834 \\
\hline
$12$ & 0.38581 & 0.38612 & 12.436 & 12.433 & 0.63200 & 0.63198 \\
\hline
$13$ & 0.37120 & 0.37145 & 14.682 & 14.625 & 0.69651 & 0.69706 \\
\hline
$14$ & 0.35809 & 0.38612 & 17.011 & 16.989 & 0.76389 & 0.76352 \\
\hline
$15$ & 0.34653 & 0.34641 & 19.477 & 19.524 & 0.82997 & 0.83131 \\
\hline
$16$ & 0.33573 & 0.33561 & 22.203 & 22.231 & 0.89980 & 0.90038 \\
\hline
$17$ & 0.32568 & 0.32573 & 25.120 & 25.111 & 0.97058 & 0.97067 \\
\hline
$18$ & 0.31677 & 0.31666 & 28.156 & 28.162 & 1.0406 & 1.0422 \\
\hline
$20$ & 0.28662 & 0.28659 & 34.610 & 34.781 & 1.1904 & 1.1885 \\
\hline
$22$ & 0.28662 & 0.28659 & 42.166 & 42.089 & 1.3404 & 1.3392 \\
\hline
$24$ & 0.27434 & 0.27438 & 50.003 & 50.085 & 1.4937 & 1.4940 \\
\hline
$26$ & 0.263647 & 0.263579 & 58.783 & 58.769 & 1.6535 & 1.6528 \\
\hline
$28$ & 0.25390 & 0.25393 & 68.260 & 68.142 & 1.8176 & 1.8152 \\
\hline
$30$ & 0.24509 & 0.24525 & 78.284 & 78.203 & 1.9782 & 1.9812 \\
\hline
\end{tabular}
\caption{\label{tab:Tab11} Thermodynamic quantities from Monte Carlo 
simulations and corresponding theoretical fits for strong disorder with 
half of the magnetic moments missing, $c = 0.5$.}
\end{center}
\vspace{-0.6cm}
\end{table}

\begin{table}
\begin{center}
\begin{tabular}{|c|c|c|c|c|c|c|}
\hline
$n$ & $m^{x=0.4}_{\mathrm{num}}$ & $m^{x=0.4}_{\mathrm{fit}}$ &
$\chi^{x=0.4}_{\mathrm{num}}$ & $\chi^{x=0.4}_{\mathrm{fit}}$ &
$\frac{d\xi}{dT}^{x = 0.4}_{\mathrm{num}}$ & $\frac{d\xi}{dT}^{x = 0.4}_{\mathrm{fit}} $   \\
\hline
$5$ & 0.63025 & 0.63036 & & & & \\
\hline
$6$ & 0.58355 & 0.58327 & & & & \\
\hline
$7$ & 0.54462 & 0.54481 & & & & \\
\hline
$8$ & 0.51280 & 0.51278 & 7.0363 & 7.0359 & 0.097086 & 0.09711\\
\hline
$9$ & 0.48570 & 0.48564 & 9.0001 & 9.0083 & 0.11011 & 0.11018 \\
\hline
$10$ & 0.46236 & 0.46230 & 11.189 & 11.188 & 0.12359 & 0.12358 \\
\hline
$11$ & 0.44190 & 0.44195 & 13.601 & 13.577 & 0.13720 & 0.13728 \\
\hline
$12$ & 0.42378 & 0.42403 & 16.158 & 16.173 & 0.15144 & 0.15128 \\
\hline
$14$ & 0.39391 & 0.39381 & 22.029 & 21.989 & 0.18022 & 0.18016 \\
\hline
$15$ & 0.38075 & 0.38091 & 25.166 & 25.208 & 0.19493  & 0.19501 \\
\hline
$16$ & 0.36932 & 0.36918 & 28.596 & 28.634 & 0.20974 & 0.21014 \\
\hline
$17$ & 0.35837 & 0.35846 & 32.245 & 32.265 & 0.22537 & 0.22554 \\
\hline
$18$ &  0.34866 & 0.34861 & 36.164 & 36.103 & 0.24193 & 0.24119 \\
\hline
$20$ & 0.33109 & 0.33112 & 44.376 & 44.394 & 0.27323 & 0.27325 \\
\hline
$22$ & 0.31595 & 0.31600 & 53.543 & 53.501 & 0.30523 & 0.30627 \\
\hline
$24$ & 0.30289 & 0.30277 & 63.347 & 63.423 & 0.33949 & 0.34023 \\
\hline
$26$ & 0.29122 & 0.29106 & 73.934 & 74.155 & 0.37480 & 0.37508 \\
\hline
$28$ & 0.28050 & 0.28060 & 85.867 & 85.693 & 0.41256 & 0.41079 \\
\hline
$30$ & 0.27112 & 0.27119 & 98.147 & 98.036 & 0.44716 & 0.44733 \\
\hline
\end{tabular}
\caption{\label{tab:Tab12}Monte Carlo data and theoretical fits 
for the magnetization, susceptibility, and $d \xi /dT$ for the 
case $c = 0.4$ of very strong disorder in the vicinity of the 
site percolation threshold, $c = 0.3116$.}
\end{center}
\vspace{-0.6cm}
\end{table}

Table~\ref{tab:Tab13} and Table~\ref{tab:Tab14} contain the self-averaging parameter $g_{2}$ for various systems sizes
for site disorder ranging from the weak regime (where $c = 0.95$), to the strongly
disordered $c = 0.40$ case in the vicinity of the percolation threshold.
A consistent feature in the dependence of $g_{2}$ on system size is an initial rise, and maximum attained for 
moderate sized systems with on the order of 700 spins.  After reaching a peak, the $g_{2}$ self-averaging parameter 
begins a steady decrease consistent with intact self-averaging.  However, the non-monotonic behavior is another  
manifestation of significant corrections to leading order scaling.

\begin{table}
\begin{center}
\begin{tabular}{|c|c|c|c|c|}
\hline
$n$ & $g_{2}^{x = 0.95}$ & $g_{2}^{x = 0.90}$ & $g_{2}^{x = 0.80}$ & $g_{2}^{x = 0.70}$ \\
\hline
$4$ & 0.011629 & 0.021031 & 0.036693 & 0.048896 \\
\hline
$5$ & 0.012967 & 0.023353 & 0.040195 & 0.052853 \\
\hline
$6$ & 0.013691 & 0.024601 & 0.041686 & 0.054397 \\
\hline
$7$ & 0.013910 & 0.025078 & 0.042294 & 0.054408 \\
\hline
$8$ & 0.014214 & 0.025350 & 0.042353 & 0.054383 \\
\hline
$9$ & 0.014231 & 0.025331 & 0.042114 & 0.053457 \\
\hline
$10$ & 0.014260 & 0.024991 & 0.041469 & 0.052606 \\
\hline
$11$ & 0.014404 & 0.025016 & 0.041237 & 0.051461 \\
\hline
$12$ & 0.014242 & 0.024941 & 0.040218 & 0.050677 \\
\hline
$13$ & 0.014139 & 0.024665 & 0.040141 & 0.049733 \\
\hline
$14$ & 0.013823 & 0.024351 & 0.039113 & 0.048602 \\
\hline
$15$ & 0.014003 & 0.024223 & 0.038562  & 0.047621 \\
\hline
$16$ & 0.013857 & 0.023888 & 0.038132 & 0.047256 \\
\hline
$17$ & 0.013882 & 0.023619 & 0.037801 & 0.046230 \\
\hline
$18$ & 0.013685 & 0.023398 & 0.037099 & 0.045487 \\
\hline
$20$ & 0.013441 & 0.023101 & 0.036060 & 0.044077 \\
\hline
$22$ & 0.013139 & 0.022324 & 0.035145 & 0.042900 \\
\hline
$24$ & 0.013017 & 0.021689 & 0.034176 & 0.041727 \\
\hline
$26$ & 0.012982 & 0.021758 & 0.033564 & 0.041068 \\
\hline
$28$ & & 0.021433 & 0.032892 & 0.039746 \\
\hline
$30$ & & & & 0.039227 \\
\hline
$32$ & & 0.020729 & 0.031422 & 0.038045 \\
\hline
$34$ & & & 0.037439 & \\
\hline
$36$ & 0.012127& 0.020170 & 0.030430 & \\
\hline
\end{tabular}
\caption{\label{tab:Tab13} Values of the self-averaging parameter $g_{2}$ for weak to moderate 
disorder strengths.} 
\end{center}
\vspace{-0.6cm}
\end{table}

\begin{table}
\begin{center}
\begin{tabular}{|c|c|c|c|}
\hline
$n$ & $g_{2}^{x = 0.6}$ & $ g_{2}^{x = 0.5}$ & $g_{2}^{x = 0.40}$  \\
\hline
$4$ &  0.057541  & 0.066634 & 0.091366 \\
\hline
$5$ & 0.061999 & 0.070882 & 0.093814 \\
\hline
$6$ & 0.063578 & 0.072695 & 0.095018 \\
\hline
$7$ & 0.062970 & 0.072895 & 0.094071 \\
\hline
$8$ & 0.062751 & 0.071787 & 0.092343 \\
\hline
$9$ & 0.061942 & 0.070998 & 0.090638 \\
\hline
$10$ & 0.060216 & 0.068958 & 0.089034 \\
\hline
$11$ & 0.059253 & 0.068091 & 0.086531 \\
\hline
$12$ & 0.058137 & 0.066394 & 0.084828 \\
\hline
$13$ & 0.057270 & 0.065569 & 0.083349 \\
\hline
$14$ & 0.055345 & 0.063845 & 0.080870 \\
\hline
$15$ & 0.054770 & 0.061972 & 0.079409 \\
\hline
$16$ & 0.053529 & 0.061077 & 0.077896 \\
\hline
$17$ & 0.052644 & 0.060114 & 0.076659 \\
\hline
$18$ & 0.052211 & 0.058970 & 0.073666 \\
\hline
$20$ & 0.050329 & 0.056281 & 0.071286 \\
\hline
$22$ & 0.048794 & 0.055587 & 0.068609 \\
\hline
$24$ & 0.047155 & 0.053513 & 0.066323 \\
\hline
$26$ & 0.046044 & 0.052511 & 0.064995 \\
\hline
$28$ & 0.045294 & 0.051496 & 0.063222 \\
\hline
$30$ & 0.044054 & 0.050261 & 0.061438 \\
\hline
$32$ & 0.043114 & 0.048754 & 0.060254 \\
\hline
$34$ & & & 0.058666 \\
\hline
$36$ & 0.041824 & 0.056226 & \\
\hline
\end{tabular}
\caption{\label{tab:Tab14} Values of the self-averaging parameter $g_{2}$ for moderate to 
strong levels of disorder.}
\end{center}
\vspace{-0.6cm}
\end{table}

\begin{acknowledgments}
Discussions with Victor Galitski and Michael Fisher are gratefully acknowledged.  Our 
numerical calculations have benefited from the University of Maryland 56 node High 
Performance Computing Cluster (HPCC).  This work has been supported by SWAN-NRI, LPS, and a University of 
Missouri Research Board Grant.

\end{acknowledgments}


\end{document}